\documentstyle[12pt]{article}
\begin{document}
\title{THE THEORY OF GRAVITATION: A TALE OF MANY QUESTIONS AND FEW ANSWERS}
\author{L. Herrera\thanks{On leave from UCV, Caracas, Venezuela, e-mail: lherrera@usal.es}\\
{Instituto Universitario de F\'isica
Fundamental y Matem\'aticas},\\ {Universidad de Salamanca, Salamanca 37007, Spain}}
\maketitle

\begin{abstract}
 We discuss on different issues pertaining the theory of gravity, which pose some unresolved fundamental questions. First we tackle the problem of observers in general relativity, with particular emphasis in tilted observers. We explain why these observers may detect dissipative processes in systems which appear isentropic to comoving observers. Next we analyze the strange relationship between vorticity and radiation, and underline the potential observational consequences of such a link. Finally we summarize all the results that have been obtained so far on the physical properties of the sources of gravitational radiation. We conclude with a list of open questions which we believe deserve further attention.
\end{abstract}

\section{Introduction}
{\it  ``There is something fascinating about science.  One gets such wholesale returns of conjecture out of such a trifling investment of fact} -- Mark Twain
\medskip

In this manuscript  I would like to elaborate on some issues that  I raised recently, at  the occasion of the ``70\&70 Gravitation Fest", which took place  in Cartagena, Colombia on September 28-30, 2016, to celebrate the 70th birthday of Rodolfo Gambini and myself. 

The text is heavily based on the  conference I delivered at that occasion, but contains some new aspects of the analyzed problems, which were absent in my conference.

I shall  focus on three  issues, recurrently appearing in the theory of gravity (in any theory of gravity, not only general relativity), which seem to me particularly appealing and endowed with potential for further development. These are:
\begin{itemize}
\item What is the role played by the observers in general relativity? What can we learn from the study of  tilted spacetimes? How observer dependent is the concept of irreversibility?
\item What is the link  between vorticity and gravitational radiation? Is it a cause--effect relationship?  Or are they concomitant? What are the possible observational consequences of such a link?
\item How to relate the gravitational radiation with the physical properties of its source? What constraints should be impossed on the source to supress the emission of gravitational radiation? What is the spacetime outside a source of gravitational radiation?  How does the emission of gravitational radiation start (as seen from the source)? How does this process cease? Are there wave tails?
\end{itemize}

\section{Tilted and comoving observers, and the definition of irreversibility (entropy)}
{\it ``Irreversibility is a consequence of the explicit introduction of ignorance into the fundamental laws.'' } M. Born
\medskip

In general relativity there exist an  ambiguity in the description of the source,  related to the arbitrariness in the choice of the four-velocity in terms of which the energy momentum tensor is split. Thus, for example,  one  possible interpretation of a given spacetime may correspond  to a congruence of observers comoving with the fluid, whereas  the other  corresponds to the  observer congruence which  has been (Lorentz) boosted, with respect to the former. In such a case,  both, the general properties of the source and the kinematic properties of the congruence would be different.
Many examples of this kind have been analyzed in the past (see for example \cite{1}--\cite{t5} and references therein)

For example, in  the case of the zero curvature FRW model, we have  a perfect fluid solution for observers at rest with respect to the timelike congruence defined by the eigenvectors of the Ricci tensor, whereas  for observers moving relative to the previously mentioned congruence of observers, it can also be interpreted as the exact solution for a viscous dissipative fluid  \cite{4}. It is worth noticing that the relative (``tilting'') velocity between the two congruences may be related to a physical phenomenon such as the observed motion of our galaxy relative to the microwave background radiation \cite{9}.

Thus,   zero curvature FRW models as described by ``tilted'' observers will detect a dissipative fluid and  energy--density inhomogeneity, as well as different values for  the expansion scalar and the shear tensor, among other differences, with respect to the ``standard''  (comoving) observers (see \cite{4} for a comprehensive discussion on this example). 

Now, the question arises: is the heat flux vector observed by tilted observers, associated to irreversible processes or not? The rationale behind such a question resides in the fact  that in the past some authors have argued that  dissipative fluids (understood as fluids whose energy--momentum tensors present a non--vanishing heat flux contribution), are not necessarily incompatible with reversible processes (e.g see \cite{14T}--\cite{16T}). 

More specifically, in the context of the standard Eckart theory \cite{17T}, a necessary condition for the compatibility of an imperfect fluid with vanishing entropy production (in the absence of bulk viscosity) is  the existence of a conformal Killing vector field (CKV) $\chi^\alpha$ such that $\chi^\alpha =\frac{V^\alpha}{T}$ where $V^\alpha$ is the four--velocity of the fluid and $T$ denotes the temperature. In the context of   causal  dissipative  theories, e.g. \cite{18T}--\cite{23T}, the existence of such CKV is also necessary for an imperfect fluid to be compatible with vanishing entropy production (see \cite{38}). 

However, such a claim should not worry us, for two reasons. On the one hand, a carefull analysis of the problem readily shows, that  the compatibility of reversible processes with  the existence of dissipative fluxes becomes trivial if a constitutive transport equation is adopted. Indeed,  in this latter case such compatibility  forces the heat flux vector to vanish as well. In other words, even if a non--vanishing heat flow vector is assumed to exist, the imposition of the CKV and the vanishing entropy production condition,  cancel the heat flux,  once a  transport equation is assumed (see \cite{NDT} for a detailed discussion on this point).
In other words, in the presence of a CKV of the kind mentioned before, the  assumption of a  transport equation whether in the context of the Eckart--Landau theory, or a causal theory, implies that  a vanishing entropy production leads to a vanishing heat flux vector. Therefore, under the conditions above,   the system is not only reversible but also non dissipative.

On the other hand,  neither LTB  \cite{25, 26, 27} nor the Szekeres spacetimes  \cite{1s, 2s} admit a CKV, accordingly we may safely conclude that the heat flux vector  appearing in these cases (at least), is associated to truly (entropy producing) dissipative processes \cite{38, 2S}.

Thus,   an  intriguing question arises, namely: how is it possible  that tilted observers may detect irreversible processes, whereas comoving observers describe an isentropic situation ?

As we shall see below, the answer to the above question is closely related to definition of entropy,  which is highly observer dependent, as illustrated, for example, by the Gibbs paradox \cite{entropy}.

Indeed, entropy  is a measure of how much is not known (uncertainty). The fact  that physical objects do not have  an intrinsic uncertainty (entropy) has been illustrated in great detail in \cite{adami, J}.

The ``antropomorphic'' nature of the concept of entropy makes itself evident in  the Gibbs paradox. In its simplest  form, the paradox appears from the consideration of a box divided by a wall in two identical parts, each of which is filled with an ideal gas (at the same pressure and temperature). Then if  the partition wall is removed, the gases of both parts of the box will mix. 

If the gases from both sides are distinguishable, the entropy of the system will rise, while there is no increase in entropy if they are identical. This leads to the striking conclusion that irreversibility (and thereby entropy), depends on the ability of the observer to distinguish, or not, the gases from both sides of the box. In other words, irreversibility would depend on our knowledge of physics \cite{bais}, confirming thereby our previous statement  that physical objects are deprived of intrinsic entropy. It can only be defined {\bf after} the number of states that are accesible  by the system, is established.

 Now, if a given physical system is studied by a congruence of comoving observers, then the three--velocity of any given fluid element is automatically assumed to vanish, whereas for the tilted observers this variable represents an additional degree of freedom. Therefore, the number of possible states in the latter case is much larger than in the former one. 

In orther to obtain  the tilted congruence (from the comoving one), we have to  submit locally Minkowskian  comoving observers to a Lorentz boost. Such  an operation, of course, is performed locally, and no global transformation exists linking both congruences. The point is that, passing from the tilted congruence to the comoving one, we usually overlook the fact  that both congruences of observers store different amounts of information. Here resides the clue to resolve the quandary mentioned above, about the presence or not of dissipative processes, depending on the congruence of observers that carry out the analysis of the system.

Thus, since  for the comoving observers  the system is dissipationless, it is clear that the increasing of entropy, when passing to the tilted congruence, should imply the presence of dissipative (entropy producing) fluxes, in the  latter.

It is instructive to take a look on this issue from a different perspective. Thus, let us  consider the transition from the tilted congruence to the comoving one.
According to the Landauer principle, \cite{lan} (also referred to as the Brillouin principle \cite{3B}--\cite{8B}), the erasure of one bit of information stored in a system requires the dissipation into the environment of a minimal
amount of energy, whose lower bound is given by 
\begin{equation}
\bigtriangleup E=kT \ln2,
\label{lan1}
\end{equation}
where $k$ is the Boltzmann constant and $T$ denotes the  temperature of the environment. 

In the theory of information, erasure, is just a reset operation restoring the system to a specific state, and is achieved by means of an external agent. In other words, one can decrease the entropy of the system by doing work on it, but then one has to increase the entropy of another system (or the environment). So to speak, Landauer principle is an expression of the fact that logical irreversibility necessarily implies thermodynamical irreversibility.

Thus, transforming to the comoving congruence, we reset the value of the three--velocity (of any fluid element) to zero, which implies that the information has been erased, and  a decrease of entropy occurs, but we have not any external agent (since we are considering self--gravitating systems), and therefore such a decrease of entropy  is accounted by the dissipative flux observed in the tilted congruence (remember that in the comoving congruence the system is dissipationless).

We may summarize the main issues addressed in this section, in the following  points:
\begin{itemize}
\item  Uncertainty (entropy) is highly dependent on the observer.
\item Comoving and tilted observers, store different amounts of information.
\item According to the Landauer principle, erasure of information is always accompanied by dissipation (you have to pay  to forgetting).
\item The quandary mentioned above, is resolved at the light of the three previous comments.
\end{itemize}

\section{Radiation and vorticity}
{\it``How often have I said to you that when you have eliminated the impossible, whatever remains, however improbable, must be the truth?''}- Sherlock Holmes in: The sign of four. 
\medskip

The first theoretical evidence relating gravitational radiation to vorticity,  appeared when it was  established  that  in an expansion of inverse powers of  $r$ (where $r$ denotes the null Bondi coordinate \cite{bondi}), the coefficient of the vorticity  at order  $1/r$ will vanish if and only if there are no news (no gravitational radiation)   \cite{1Rv}, implying that a frame dragging effect is associated with gravitational radiation. This result was further confirmed in \cite{2Rv}--\cite{5Rv}.

The intriguing fact is that these two (very different) phenomena are related. More specifically, we wonder if there is any physical reason behind  the link between vorticity and radiation.

The hint to solve this quandary comes from an idea put forward by  Bonnor in order to explain the appearance of vorticity in the spacetime generated by a charged magnetic dipole \cite{8Rv}. 

For such a system, as was observed by Bonnor, there exists  a non--vanishing component of the Poynting vector, describing a flow of electromagnetic energy round in circles \cite{9Rv}. He then suggested that such a circular flow of energy affects inertial frames by producing vorticity of congruences of particles, relative to the compass of inertia. This conjecture has been  shown to be valid for a general axially symmetric stationary electrovacuum metric \cite{10Rv}. It is worth mentioning that this ``circular'' flow of electromagnetic energy is endowed with a solid physical meaning, since it is absolutely necessary  in order to preserve the conservation of angular momentum, in the well known  ``paradox'' of the rotating disk with charges and a solenoid \cite{9Rv}.

Then it came to our minds  \cite{4Rv}, the idea that a similar mechanism might be at the origin  of vorticity in the gravitational case, i.e.  that a circular flow of gravitational energy would produce  vorticity. However the nonexistence of   a local and invariant  definition of  gravitational energy, rose at that time the question about what expression for the ``gravitational'' Poynting vector  should be used.   In \cite{5Rv} we tried with the super--Poynting vector based on the Bel--Robinson tensor \cite{11Rv}--\cite{14Rv}.

Doing so, we succeded to establish the link between  gravitational radiation and vorticity, invoking   a mechanism similar to that proposed by Bonnor for the charged magnetic dipole. It was shown  that the resulting vorticity is always associated to a circular flow of superenergy on the plane orthogonal to the vorticity vector. It was later shown that the vorticity appearing in stationary vacuum spacetimes also depends on  the existence of a flow of superenergy on the plane orthogonal to the vorticity vector \cite{7Rv}. Furthermore in \cite{16Rv} it was shown that not only gravitational but also electromagnetic radiation produces vorticity. In this latter case we were able to isolate contributions from, both, the electromagnetic Poynting vector as well as from the super--Poynting vector.

It is worth noticing that for unbounded configurations  (e.g. cylindrically symmetric sytems), radiation  does not produce vorticity if there are no   circular flows of superenergy on any  plane of the $3$-space (e.g. Einstein--Rosen spacetime), which reinforces further the role of the circular flow of super--energy as the link between radiation and vorticity.

Besides the evident theoretical interest of the issue discused in this section, we would like to emphasize the very important potential observational consequences of the association of  vorticity with radiation. Indeed, the   direct experimental evidence of the existence of the Lense--Thirring effect \cite{22Rv}--\cite{ciu2}  brings out the high degree of development achieved in  the   technology required to measure rotations. In the same direction point recent proposals to detect frame dragging by means of ring lasers \cite{rl1}--\cite{rl4}. Also it is worth mentioning the possible use of atom interferometers \cite{at1}-- \cite{at3}, atom lasers  \cite{atl},  anomalous spin--precession experiments \cite{sp} and matter wave Sagnac interferometers \cite{CH},  to measure vorticity.

Le us close this section with a remark about an important feature. 

At order ($1/r^2$) there are contributions to the vorticity with a time dependent term not involving news (i.e. not associated with gravitational radiation). This last term represents the class of non-radiative motions discussed by Bondi \cite{bondi} and may be thought to correspond to the tail of the wave, appearing after the radiation process has ended \cite{BoNP}. Thus, the obtained expression allows for ``measuring'' (in a gedanken experiment, at least) the wave-tail field. This in turn implies that observing the gyroscope, for a period of time from an initial static situation until after the vanishing of the news, should allow for an unambiguous identification of a gravitational radiation process.

However, as it has been recently shown (see the subsection 4.6 below)  the transition from a state of radiation (gravitational) to an equilibrium state, is not forbidden, after a small time interval of the order of magnitude of the hydrostatic time, the relaxation time, and the thermal adjustment time.
This result is in contradiction with previous results \cite{bondi}, \cite{BoNP}--\cite{MBF}, that suggest that such a transition is forbidden, due to the appearance of the wave tails mentioned above.

Therefore the detection (or not) of the wave tail, would have very important theoretical consequences.

\section{Gravitational radiation and its source}
{\it ``I have no data yet. It is a capital mistake to theorize before one has data. Insensibly one begins to twist facts to suit theories, instead of theories to suit facts''}.--Sherlock Holmes in: A scandal in Bohemia.

\medskip

In the sixties, there was a blossom of very  powerful methods to study gravitational radiation, beyond the well known linear approximation, \cite{bondi, 8rs, NP, BR}. All of them focused on the behaviour of the field very far from the source, whose specific properties did not enter into the discussion, to avoid the appearance of caustics and similar pathologies.  

Besides the many fundamental results obtained from these methods, their main merit consists in including non--linear effects, which are known to play a very important role in general relativity. 
The absence of a detailed discussion on the behaviour of the sources of gravitational radiation, was the triggering motivation to undertake the task to develop a general formalism that could provide a description of the effects of gravitational radiation on the physical properties of the source, and viceversa.

In other words we searched   to establish the relationship between gravitational radiation and source properties.

The fact that we want to describe gravitational radiation, force us  to depart from  the spherical symmetry.  On the other hand, since we are willing to provide an analytical description of the problem, avoiding as much as possible to resort to numerical methods, we need to impose additional restrictions. Thus, we shall rule out cylindrical symmetry  on physical grounds, but we shall assume  axial and reflection symmetry, which as shown  in \cite{5bis}  is the highest degree of symmetry  of the Bondi  metric \cite{bondi}, which does  not prevent the emission of gravitational radiation.

The general formalism mentioned before was presented in  \cite{Ref1}. It was obtained by using the $1+3$ formalism \cite{ref6, ref7, ref8}, in a given coordinate system, and  resorting to a set of scalar functions known as Structure Scalars \cite{ref9}, which have been shown to be very useful in the description of self--gravitating systems \cite{ref10}--\cite{18b}.

A detailed description of this general formalism may be found in \cite{Ref1}, here we shall focus on the main results obtained so far on this issue, and the open questions which we believe deserve further attention.

\subsection{The shear--free case}
We started by analyzying the shear free case  \cite{ref3}. Under such a condition we have found that
 for a general dissipative and anisotropic fluid, vanishing vorticity, is a  necessary and sufficient condition for the magnetic part of the Weyl tensor  to vanish, thereby providing a generalization of  the same result for perfect fluids obtained in \cite{b1, b2, glass}.
This result, in turn, implies  that  vorticity should necessarily appear if the system  radiates gravitationally. We stress that this result is not restricted to the axially (and reflection) symmetric case. This further reinforces  the well established  link between radiation and vorticity discussed in the previous section.

If besides the shear--free condition, we assume further that the fluid is geodesic, then  the vorticity  vanishes (and thereoff  the magnetic part of the Weyl tensor) and  no gravitational radiation is produced. In this case our study generalizes the models studied by Coley and McMannus \cite{c1, c2}. 
A similar result is obtained for the cylindrically symmetric case \cite{cyl},  suggesting a link between the shear of the source and the generation of gravitational radiation.

Finally, in the geodesic case, we also observe that,  in the non-dissipative case, the models do not need to be FRW (as already stressed in \cite{c1}), however  the system tends to the FRW spacetime  (if $\Theta$ is positive). In presence of  dissipative fluxes, such a tendency does not appear, which illustrates further how relevant, the dissipative processes may be. 

\subsection{The perfect, geodesic fluid}
Next we have considered the the restricted case, when  the fluid is  perfect and geodesic, without assuming {\it ab initio} the shear--free condition \cite{ref2}.
As the result of such study we have found that:  All possible models compatible with our metric, and the  perfect fluid plus the geodesic condition, are FRW, and accordingly non--radiating (gravitationally). It seems that, both, the geodesic and the non--dissipative conditions, are quite restrictive, when looking  for a source  of gravitational waves. 

It is worth noticing that, not only in the case of dust, but also in the absence of dissipation in a perfect fluid, the system is not expected  to radiate (gravitationally) due to the reversibility of the equation of state. Indeed,  radiation is an irreversible process, a fact that emerges at once  if  absorption is taken into account and/or Sommerfeld type conditions, which eliminate inward traveling waves, are imposed. Therefore,  the irreversibility of the  process of emission of gravitational waves, must be reflected in the equation of state through an entropy increasing  (dissipative) factor.

We should recall that geodesic fluids not belonging to the class considered here (Szekeres) have also been shown not to produce gravitational radiation. This strengthens further the case of the non--radiative character of pure dust distributions.

\subsection{The dissipative, geodesic fluid}
Since dissipation should be present in any scenario where gravitational radition is produced, we decided to study the simplest possible fluid distribution which we might expect to be  compatible with a gravitational radiation, i.e. a dissipative dust under the geodesic condition \cite{ref4}.

Two  possible subcases were considered separately, namely: the fluid distribution is assumed, from the beginning, to be vorticity--free, or not. 

In the former case, it was shown that the vanishing of the vorticity implies the vanishing of the heat flux vector, and therefore, the resulting spacetime is FRW.

In the latter case, it is shown that the enforcement of the regularity conditions  at the center, implies the vanishing of the dissipative flux, leading also to a FRW spacetime.

Thus  all possible models, sourced by a dissipative  geodesic dust fluid, belonging to the family of the line element considered here, do not radiate gravitational waves during their evolution, unless regularity conditions at the center of the distribution are relaxed.  

In other words physically acceptable models require the inclusion of, both, dissipative and anisotropic stresses terms, i.e. the geodesic condition must be abandoned. In this case, purely analytical methods are unlikely  to be sufficient to arrive at a full description of the source, and one has to resort to numerical methods.

\subsection{The space--time outside the source of gravitational radiation}
Based on the fact that  the process of  gravitational radiation is an irreversible one, and therefore must entail dissipative processes within the source, we should conclude that  there should be an  incoherent radiation (null fluid)  at the outside of the source, produced by those dissipative processes. Keeping this fact in mind, we should remark that the Bondi--Sachs metric \cite{bondi}, \cite{8rs}, should be regarded as an approximation to the space--time outside the source, when the null fluid produced by the dissipative processes is neglected.

 Starting with the description of this null fluid, we apply the formalism developped in \cite{Ref1}, to study some of the properties of such a null fluid \cite{gva}. 

As the main result of our study we found that the absence of vorticity implies that the exterior spacetime is either static or spherically symmetric (Vaidya). Reinforcing thereby the fundamental role of vorticity in any process involving production of gravitational radiation, already stressed. 

The spherically symmetric case (Vaidya) was, asymptotically, recovered within the context of the 1+3 formalism \cite{gva}.
\subsection{Leaving the equilibrium}
Next, as an application of our general method \cite{Ref1}, we analyzed the situation, just after its departure from hydrostatic and thermal equilibrium, at the smallest time 
scale at which the first signs of dynamic evolution appear  \cite{oeq}. Such a time scale is smaller than the thermal relaxation time, the thermal adjustment time and the hydrostatic time.
Specifically we were able to answer to the following questions:
1) what are the first signs of non--equilibrium?
2) which physical variables do exhibit such signs?
3) what does control the onset of the dynamic regime, from an equilibrium initial configuration?

 It was obtained  that the onset of non--equilibrium will critically depend on a single function directly related to the time derivative of the vorticity. Among all fluid variables (at the time scale under consideration), only the tetrad component of the anisotropic tensor, in the subspace orthogonal to the four--velocity and the Killing vector of axial symmetry, shows signs of dynamic evolution. Also, the first step towards a dissipative regime begins with a non--vanishing time derivative of the heat flux component along the meridional direction. The magnetic part of the Weyl tensor vanishes (not so its time derivative), indicating that the emission of gravitational radiation will occur at later times. Finally, the decreasing of the effective inertial mass density, associated to thermal effects, was clearly illustrated \cite{oeq}.

\subsection{Reaching the equilibrium}
Next, we   described the transition of a gravitationally radiating, axially and reflection symmetric dissipative fluid, to a non--radiating state  \cite{re}. What we wanted to elucidate was if, very shortly after the end of the radiating regime, at a time scale of the order of the thermal relaxation time, the thermal adjustment time and the hydrostatic time (whichever is larger), the system reaches the equilibrium state. We recall that in  all the studies carried out in the past, on gravitational radiation outside the source, such a transition to a static case, is forbidden. However, neither  of these studies include the physical properties of the source, giving rise the possibility of a  quite different result.

Using the general method presented in \cite{Ref1},  we were able to prove that the system does in fact reach the equilibrium state. This result is at variance with previous results mentioned above, implying that such a transition is forbidden. The reason for such a discrepancy resides in the fact that some elementary, but essential, physical properties of the source, have been overlooked in these latter studies.
Our result strengths further the relevance of the physical properties of the source, in any discussion about the physical properties of the field. Also, it emphasizes the need to resort to global solutions, whenever important aspects about the behaviour of the gravitational field are discussed. In other words, the coupling between the source and the external field may introduce important constraints on the physical behaviour of the system, implying that details of the source fluid cannot be left out, because they may be relevant to distant GW scattering.

\subsection{The quasi--static regime}
As it is well known, in the study of self--gravitating fluids we may consider three different possible regimes of evolution, namely: the static, the quasi--static and the dynamic.

In the static   case, the spacetime admits a timelike, hypersurface orthogonal, Killing vector. Thus, a coordinate system can always be choosen, such that all metric and physical variables are independent on the time like coordinate. The static case, for axially and reflection symmetric spacetimes, was studied in \cite{static}.

Next, we have the full dynamic case where the system is considered to be out of equilibrium (thermal and dynamic), the general formalism to analyze this situation, for axially  and reflection symmetric spacetimes was developed in \cite{Ref1}.

In between the two regimes described above, we have the quasi--static evolution.

In this regime the system is assumed to evolve, although sufficiently slow, so that it can be considered to be in equilibrium at each moment. This means that the system changes slowly, on a time scale that is very long compared to the typical time in which the fluid reacts to a slight perturbation of hydrostatic
equilibrium. This typical time scale is called hydrostatic time scale (sometimes this time scale is also referred to as dynamical time scale). 

Thus, in this regime  the system is always very close to  hydrostatic equilibrium and its evolution may be described as a sequence of equilibrium configurations. 

In other words, we may assume safely the quasi--static approximation (QSA) whenever all the relevant characteristic times of the system under consideration, are much larger than the hydrostatic  time.

This assumption is very sensible because
the hydrostatic time scale is very small for many phases of the life of the star.
It is of the order of $27$ minutes for the Sun, $4.5$ seconds for a white dwarf
and $10^{-4}$ seconds for a neutron star of one solar mass and $10$ Km radius.
It is a fact  that any of the stellar configurations mentioned above, generally (but not always), 
changes on a time scale that is very long compared to their respective
hydrostatic time scales. 

Motivated by the comments above, we applied  the  framework developped in \cite{Ref1},  to carry out a study of axially  and reflection symmetric fluids in the quasi--static regime \cite{ce}. 

For doing that we needed to introduce different invariantly defined ``velocities'', in terms of which the QSA is expressed.

It was obtained that  the shear and the vorticity of the fluid, as well as the dissipative fluxes, may affect the (slow) evolution of the configuration, as to produce ``splittings'' within the fluid distribution.

Also it was shown that  in the QSA, the contributions of the gravitational radiation to the components of the super--Poynting vector do not necessarily vanish. However,  it appears   that if at any given time,
the magnetic part of the Weyl tensor vanishes, then it vanishes at any other time afterwards. Thus it is not reasonable to expect gravitational radiation from a physically meaningful system, radiating for a finite period of time (in a given time interval) in the QSA.

\section{OPEN ISSUES}
As the reader should easily understand, the three issues considered here, still present a great deal of unanswered  questions.  Below we display a partial list of problems which we believe deserve some attention:

\begin{itemize}

\item How could one describe the ``cracking'' (splitting) of the configurations, in the context of this formalism ? How the appearance of such crackings would affect the emission of gravitational radiation?
\item We do not have  an exact solution (written down in closed analytical form)  describing gravitational radiation in vacuum, from bounded sources.  Accordingly, any specific modeling of  a source, and its matching to an exterior, should be done numerically.  How such a matching could be done?
\item It should be useful to introduce the concept of the  mass function, similar to the one existing in the spherically symmetric case (an extension of the Bondi mass for the interior of the source). This could be relevant, in particular, in the  matching of the source to a specific exterior. 
\item How does the system analyzed in \cite{Ref1}  look for tilted observers? Is it possible that tilted observers detect gravitational radiation from systems that are  non--radiating (gravitationally), as seen by comoving observers? 
\item What could we learn by imposing different kind of symmetries (more general that isometries) on the axially symmetric dissipative fluids studied in \cite{Ref1}.
\item What is the threshold of sensitivity in the measure of vorticity, reached by the present technology? What are  the expected values for the potential future developments of different experiments aiming the detection of rotations? Are these values within the range expected for   realistic sources of gravitational radiation?
\end{itemize}

\section{Acknowledgments}
It is a real pleasure to thanks the organizying committee of  ``70\&70 Gravitation Fest'': Antonio C. Guti\'errez-Pi\~neres (Universidad Tecnol\'ogica de Bol\'\i var); Edison Montoya (Universidad Industrial de Santander); Jorge Mu\~niz (Universidad Tecnol\'ogica de Bol\'\i var) and Luis A. N\'u\~nez (Universidad Industrial de Santander), for their generosity and the great effort deployed to make the event possible.

\end{document}